\documentclass[%
 reprint,
 superscriptaddress,
 amsmath,amssymb,
 aps,
]{revtex4-2}
\usepackage[dvipsnames]{xcolor}
\usepackage{soul}
\usepackage{graphicx}
\usepackage{dcolumn}
\usepackage{bm}
\usepackage{natbib}
\usepackage{hyperref} 
\usepackage{ulem} 

\begin{document}

\preprint{APS/123-QED}

\title{
Strain-induced splitting of the CCDW-NCCDW phase transition in 1T-TaS$_2$}

\author{M. M. Tyumentsev}

\affiliation{Institute of Radioengineering and Electronics of RAS, 11 Mokhovaya, 125009 Moscow, Russia}
\affiliation{Physics Department, HSE University, 20 Myasnitskaya Ulitsa, 101000 Moscow, Russia} 
\author{V. E. Minakova}
\affiliation{Institute of Radioengineering and Electronics of RAS, 11 Mokhovaya, 125009 Moscow, Russia}

\author{N. I. Fedotov}
\affiliation{Institute of Radioengineering and Electronics of RAS, 11 Mokhovaya, 125009 Moscow, Russia}
\affiliation{Physics Department, HSE University, 20 Myasnitskaya Ulitsa, 101000 Moscow, Russia}

\author{S. V. Zaitsev-Zotov}
\email{serzz@cplire.ru}
\affiliation{Institute of Radioengineering and Electronics of RAS, 11 Mokhovaya, 125009 Moscow, Russia}
\affiliation{Physics Department, HSE University, 20 Myasnitskaya Ulitsa, 101000 Moscow, Russia}

\date{\today}

\begin{abstract}
The effects of uniaxial and biaxial tensile strain on the $\rho_{xx}$ and $\rho_{yy}$ components of the resistivity tensor and the commensurate --- nearly commensurate CDW (CCDW-NCCDW) transition temperature in 1T-TaS$_2$ are studied. At room temperature, uniaxial tensile strain increases the resistivity tensor components by a comparable magnitude both parallel and perpendicular to the strain axis. In the case of biaxial strain, up to 20~K decrease in the CCDW-NCCDW phase transition temperature is observed. In the case of uniaxial strain, a new phase with two different CCDW-NCCDW phase transition temperatures is observed, the splitting exceeds 10 K. The occurrence of such a phase is associated with the transition of the CDW into the commensurate state along the tensile strain direction while maintaining nearly commensurability along the perpendicular one. The results allow to justify various models widely used in analysis of transport properties of 1T-TaS$_2$ in commensurate and nearly commensurate states. 

\end{abstract}

\maketitle

\section{\label{sec:level1}Introduction}
Charge density waves (CDWs) constitute one of the most fascinating manifestations of collective electronic order in low-dimensional systems (as recent reviews see \cite{radic2025charge,pouget2024structural}). These periodic modulations of the electronic charge density, intrinsically coupled to concomitant lattice distortions, emerge from the interplay between electron-electron correlations and electron-phonon interactions. Layered materials hosting CDW order have attracted sustained interest not only for their rich phase diagrams but also because the charge-ordered state is often intimately linked to other correlated ground states, such as superconductivity, ferro- and antiferromagnetism, Mott physics and spin liquid \cite{neto2001charge,yumigeta2021advances,law20171t}. 

Transition metal dichalcogenides (TMDs) with the general composition MX$_2$ (here M = Ta, Nb, V and X = S, Se, Te) is a large group of layered materials where the CDW competes with superconductivity and Mott state  \cite{neto2001charge,manzeli20172d}. 
Among TMDs, 1T-TaS$_2$ stands out as a correlated electron system, exhibiting a rich hierarchy of charge density wave phases driven by intricate interplay between electron-lattice coupling, electronic correlations, and disorder \cite{rossnagel2011origin}. During cooling, a cascade of phase transitions of the CDW occurs.  
At temperatures of 353-355~K and 180-220~K, two hysteretic transitions occur into a nearly commensurate CDW (NCCDW) phase and commensurate CDW (CCDW) phase, respectively \cite{Wilson01122001,scruby1975role,bayliss1984thermal}. In the CCDW phase, a $\sqrt{13} \times \sqrt{13}$ superstructure is formed that evenly covers the surface of the crystal \cite{Wilson01122001,scruby1975role}. Tantalum atoms form clusters in which 12 atoms are shifted towards the central one, forming a Star of David. In the NCCDW phase, the clusters form dielectric superclusters with hexagonal orientation, at the boundaries of which the CDW gap is suppressed. 
The sensitivity of the electronic ground state ranging from metal to Mott insulator to subtle changes in the CDW periodicity makes 1T-TaS$_2$ a fertile platform for exploring phase competition and non-equilibrium quantum phenomena \cite{demsar2002femtosecond,sipos2008mott,stojchevska2014ultrafast}.

These transitions result from the temperature dependence of the CDW wave vector, which decreases by about 3\% from $0.286a^*$ to $0.277a^*$ with a decrease in temperature from 331~K to 190~K \cite{scruby1975role}, where $a^*$ is the
matrix reciprocal lattice vector.  The phase transition NCCDW-CCDW with $\sim 1$\% \ step-like jump of the CDW wave vector is accompanied by the jump of the crystal lattice constant by $\sim 0.7\cdot 10^{-3}$ \cite{svetin2014transitions}. This direct coupling between the CDW periodicity and the host lattice parameters implies that external strain should serve as a potent, continuous tuning knob to control the phase stability, transition temperatures, and domain textures. 

Strain engineering has emerged as a powerful and versatile strategy for controlling the structural and electronic properties of two-dimensional materials \cite{roldan2015strain,dai2019strain,qi2023recent,zhang2025tunable}. In the context of CDW systems, strain constitutes a natural tool, as the charge order itself is tied to a lattice distortion and thus responds sensitively to mechanical perturbations.

Previous attempts to strain-engineer 1T-TaS$_2$ relied on differential thermal expansion with rigid substrates, enabling biaxial strain up to ~0.1$\%$ \cite{svetin2014transitions, zhao2017tuning}. While these studies demonstrated shifts in transition temperatures, this approach offers only discrete, fixed strain states, prevents in-situ tuning, and is fundamentally limited to biaxial deformation.
Application of uniaxial strain is another way to control electronic states in 1T-TaS$_2$ \cite{nicholson2024gap, luque2025strain}. In Ref.~\cite{nicholson2024gap}, uniaxial strain for controlling correlated behavior via interlayer coupling in layered materials was used in ARPES and X-ray diffraction studies. Gap collapse and flat band induced by uniaxial strain in 1T-TaS$_2$ have been reported  below the commensurate charge density wave (CCDW) transition temperature. 
Detailed studies of the effects of  both uniaxial and biaxial strain on electronic transport properties and transition temperatures, to the best of our knowledge, are still lacking.

In this work, we implement a biaxial tensile device that enables in-situ application of continuous, tunable uniaxial and biaxial tensile strain to 1T-TaS$_2$ flakes across a substantially expanded range \cite{ZZ_compact_comp}. This technique makes it possible to obtain the dependencies of the transition temperatures between different phases of the CDW on uniaxial and biaxial deformations. We systematically investigate how uniaxial and biaxial deformation modifies the hysteretic transitions, the stability regions of the NCCDW and CCDW phases, and the associated transport properties. Our main conclusion is that uniaxial deformation can lead to the splitting of the CCDW-NCCDW transition with respect to the strain direction. The emergent resistivity anisotropy, absent under biaxial strain, suggests the formation of a strain-induced electronic phase with lowered symmetry, potentially involving aligned CDW domains (striped CDW phase). Our results reveal a significant modulation of the CCDW-NCCDW transition, providing new insights into the strain sensitivity of the CDW phase diagram in this prototypical TMD. These findings not only deepen our understanding of the coupling between charge order and lattice degrees of freedom in 1T-TaS$_2$ but also establish strain engineering as an effective route for tuning correlated electronic phases in layered materials for potential device applications.  
Our results establish anisotropic strain as a powerful new tool for controlling correlated states in 1T-TaS$_2$ and reveal its unique role in unlocking directional quantum phenomena. 

\section{\label{sec:level1}Experimental methods} 
\begin{figure}
\includegraphics[width=1\linewidth]{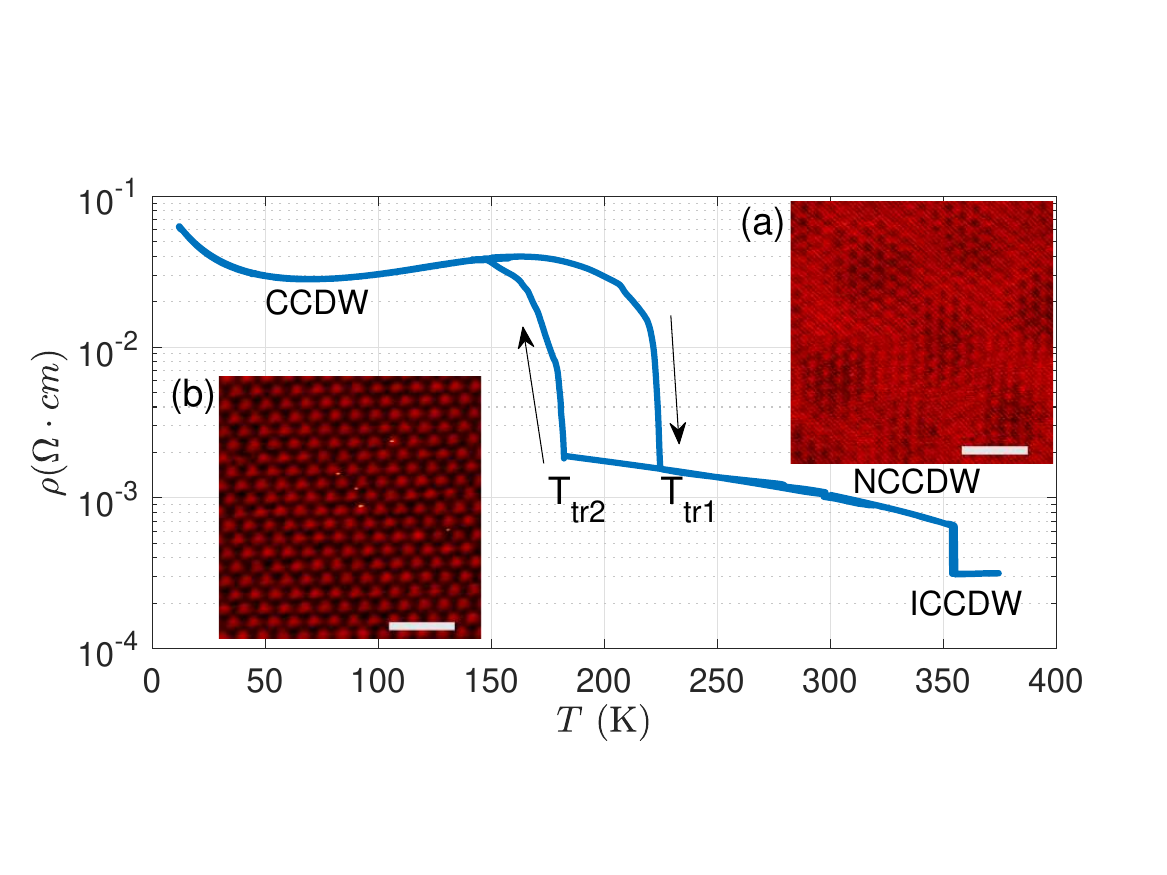}
\caption{\label{fig:R(T)} Temperature dependence of the resistivity of a bulk 1T-TaS$_2$ crystal. The arrows show directions of the measurements. Insets show STM images in (a) NCCDW ($T=300$~K) and (b) CCDW ($T=78$~K) states. Scale bars are 5 nm.}
\end{figure}

1T-TaS$_2$ crystals were synthesized by the iodine transport method \cite{ravy2012high,geremew2019bias}. An evacuated and sealed ampule containing stoichiometric amounts of Ta and S, along with a small amount (5 mg/cm$^3$) of iodine, was maintained in a two-zone furnace at a temperature gradient of 820-920°C for three weeks. At the end of the synthesis, quenching was performed by rapidly immersing the ampule from the furnace into water. As a result, mirror-like plate-shaped hexagonal faced crystals of  1T-TaS$_2$ with lateral sizes of up to 1 cm were grown. Transport measurements and ultra high vacuum scanning-tunneling microscopy (STM) studies reveal known properties of 1T-TaS$_2$ (Fig.\ref{fig:R(T)}).

\begin{figure}[h]
\includegraphics[height=2.8cm]{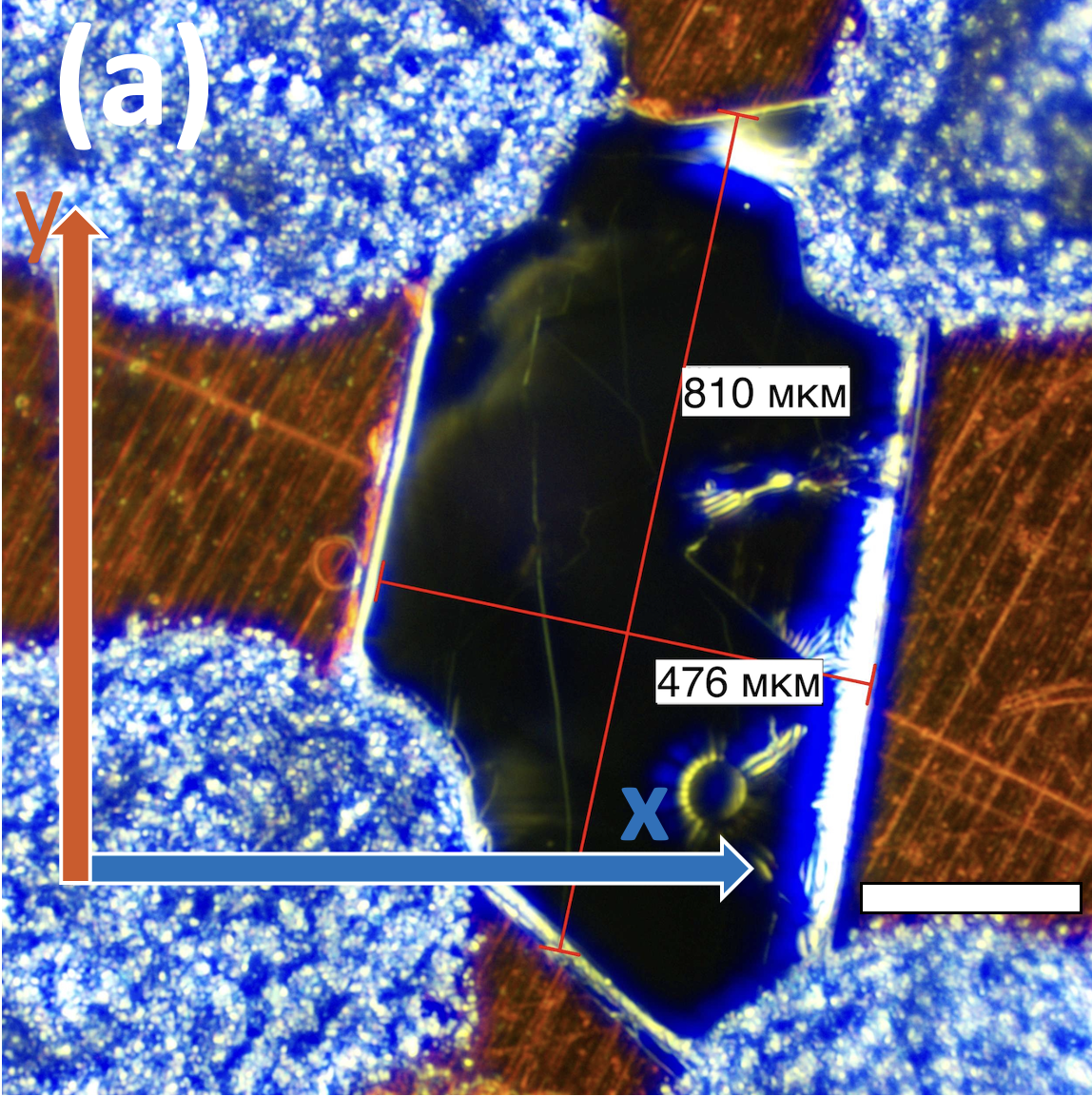} \includegraphics[height=2.8cm]{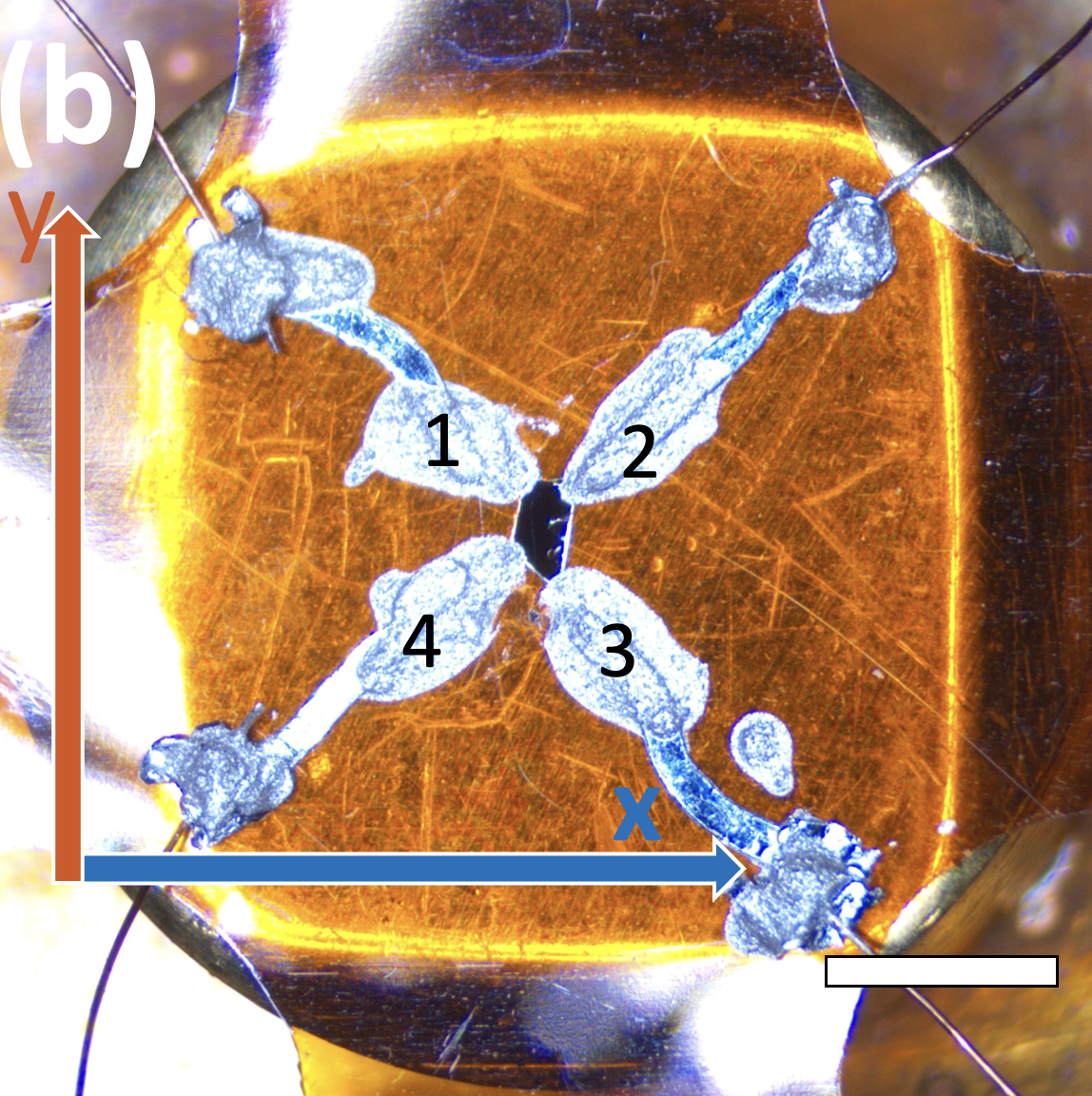} \includegraphics[height=2.8cm]{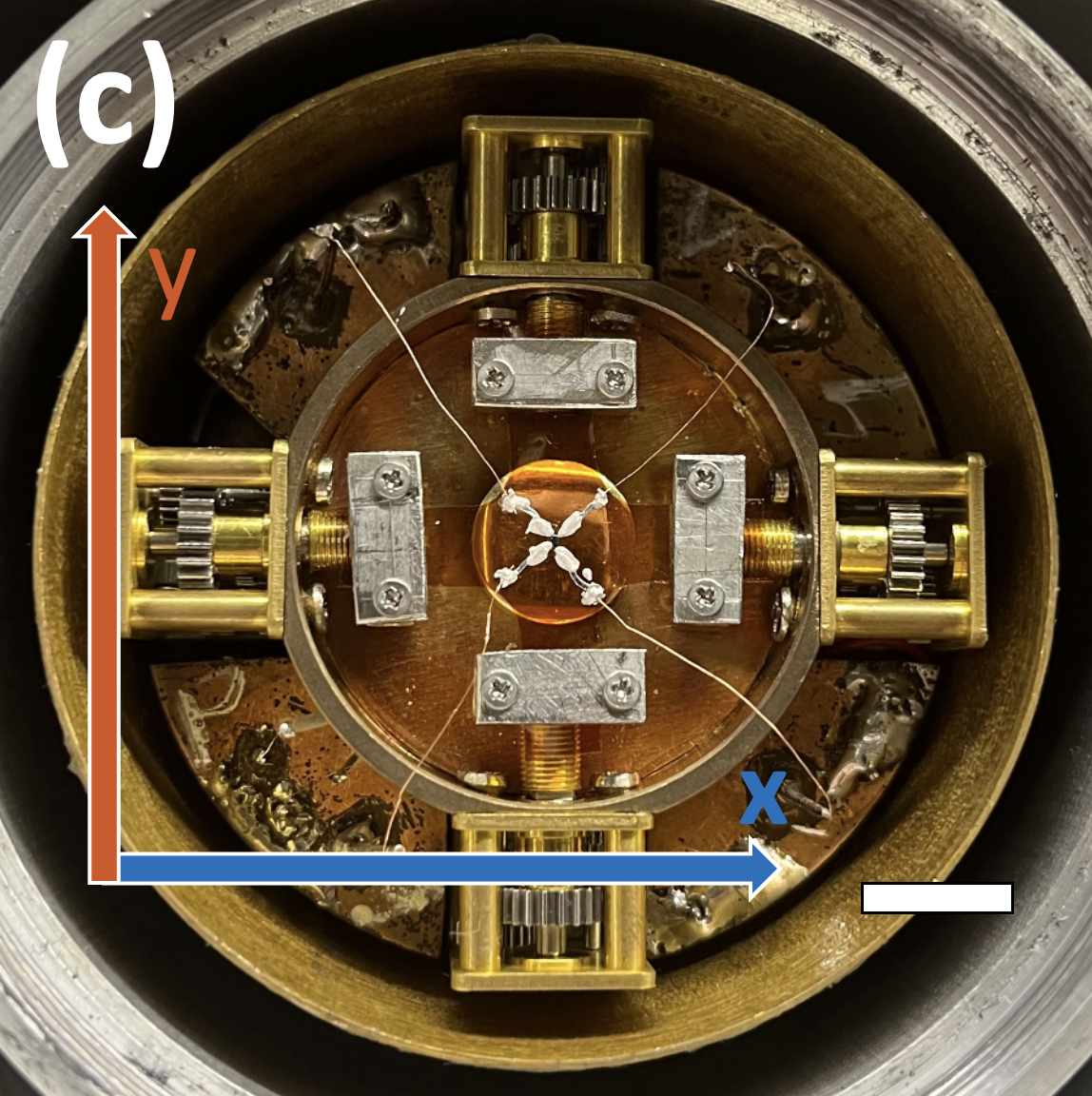}
\caption{\label{fig:samp} Photos of the sample \# 2 mounted on the biaxial tensile strain device at three different magnifications: a) The  sample with contacts (scale bar is 200~$\mu$m); b) Polyimide cross with the sample on the heat exchange finger; contact numerations is shown by numbers (scale bar is 2~mm); c) Polyimide cross with the sample mounted in the tensile strain device (scale bar is 1~cm).}
\end{figure}

We used a tensile strain device described in Ref.~\cite{ZZ_compact_comp} for the experimental study of the effect of uniaxial and biaxial strain on the transport properties of 1T-TaS$_2$ in a wide temperature range. A freshly cleaved flake of 1T-TaS$_2$ with lateral sizes of a few hundred micrometers was glued to the center of a cross-like substrate made of 75 $\mu$m thick polyimide film using a thin layer of epoxy resin. The upper layers of the crystal were carefully removed using adhesive tape in  order to achieve the desired thickness.  Electrical contacts were made using silver paint. A photo of a typical sample is shown in Fig.~\ref{fig:samp} (a). 

The cross-like substrate with a sample was mounted in the biaxial tensile strain device \cite{ZZ_compact_comp} (Fig.~\ref{fig:samp} (c)), enabling independent strain application along two orthogonal axes. The nominal strain of the substrate $\varepsilon_{x}, \varepsilon_{y}$, were determined by counting the steps of the stepper motor. The actual strain of the sample is less than the nominal strain of the substrate. The device is made of brass. Brass and polyimide have approximately the same coefficient of thermal expansion, which is slightly less than the coefficient of thermal expansion of 1T-TaS$_2$. In addition, upon cooling, 1T-TaS$_2$ crystals are subjected to abrupt compression at the moment of transition to a CCDW phase and abrupt elongation at the moment of transition to an NCCDW phase \cite{sezerman1980thermal}. This results in an additional compression of about 0.1\%. The strains discussed in this paper are carried out against the background of an isotropically compressed crystal.

Electrical measurements were made using the 4-probe method in the van der Pauw geometry in the current-controlled regime. We use the following notation: $R_x =V_{43}/I_{12}$, $R_y=V_{23}/I_{14}$, where $I_{i,j}$ is the current applied to contacts $i$ and $j$, and  $V_{i,j}$ is the respective voltage (see Fig.~\ref{fig:samp}(b) for contact numeration). The resistances $R_x$ amd $R_y$ were measured sequentially using the commutation scheme described in Ref.~\cite{ZZ_compact_comp}. The sheet resistance was calculated using the van der Pauw method \cite{philips1958method}. The thickness was calculated based on the assumption that the bulk resistivity value is $9\cdot 10^{-4} {\ } \Omega \cdot$cm \cite{THOMPSON1971981} and varied in the range 30-300~nm for the samples studied.
With such a crystal thickness, the phase diagram coincides with that of a bulk crystal \cite{yoshida2014controlling}. Additionally, it has been found that such a crystal thickness allows for the transfer of a reasonably large fraction of the substrate strain to the samples. 

Scanning tunneling microscopy (STM) study was performed under ultrahigh vacuum conditions (the base vacuum is $2\times 10^{-12}$~Torr) by using Omicron LT SPM. We use silver epoxy EPO-TEX\textregistered \ H20E to glue a sample to a flexible copper foil  stripe fixed to the Omicron flag style sample holder.  Atomically clean surface was obtained by sample cleavage {\it in situ}. Sample strain was obtained by bending the stripe {\it in vacuo} at room temperature. The nominal value of the strain was calculated from the turn angle of the bending screw.

The results of this paper are based on investigations of almost two dozen samples with varying shapes and thicknesses. The results described below were obtained in two representative samples with the following parameters. Sample \#1: $R_x = 5.51 { \ }\Omega$, $R_y = 11.2{ \ }\Omega$, sample \#2: $R_x = 2.34 { \ }\Omega$, $R_y = 86.45 { \ }\Omega$. 

\section{\label{sec:level1}Experimental results}

In the initial state, the temperature dependence of the resistivity of the studied samples coincides with that reported earlier \cite{THOMPSON1971981}: transition temperatures, heights, and hysteresis loop widths are typical for 1T-TaS$_2$ (Fig.~\ref{fig:R(T)}).

\begin{figure}[h]
\includegraphics[width=1\linewidth]{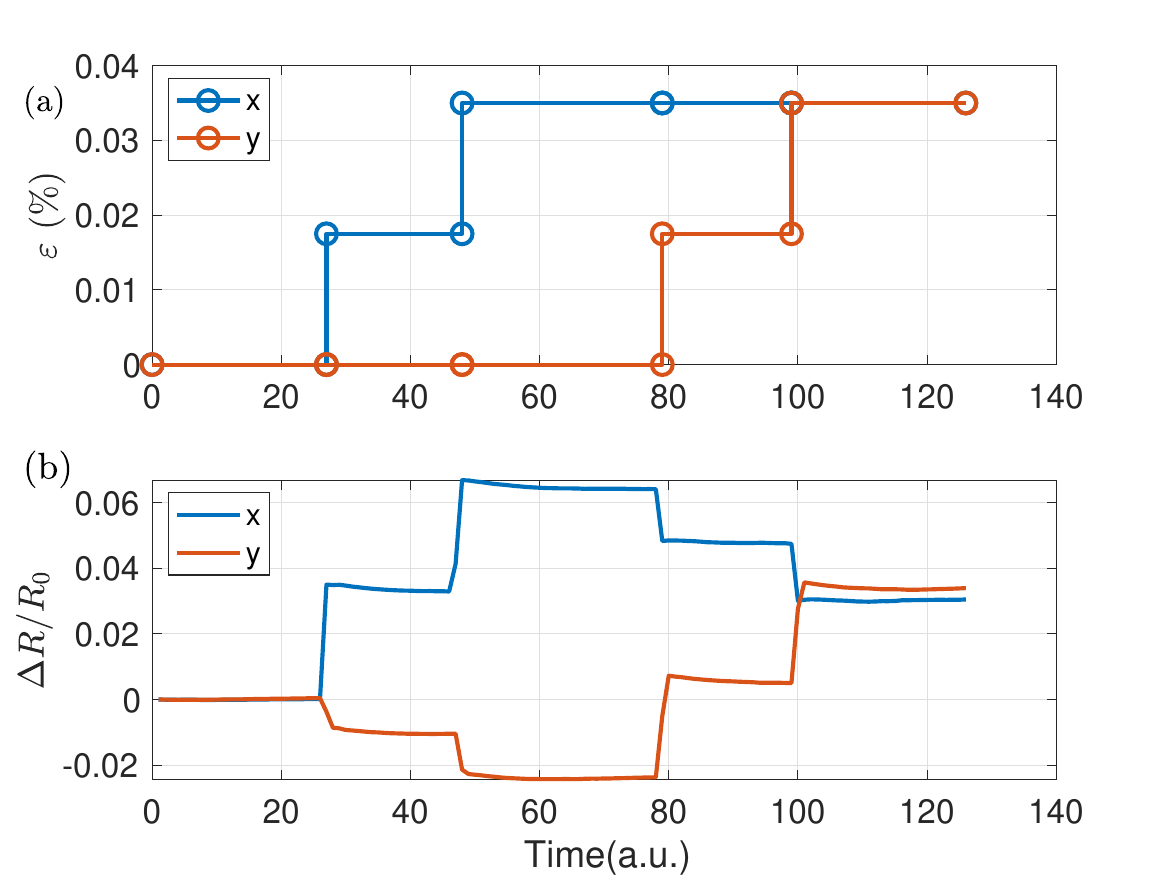}
\caption{\label{fig:str} a) Step-like deformation applied sequentially along two axes; b) respective dependence of the resistances $R_x$, $R_y$  upon a step-like deformation change, measured in sample \#1.  Sample thickness = 280 nm.  $T=300$~K.}
\end{figure}

In the first stage, the dependence of the sample resistance on the uniaxial and biaxial strain at room temperature was studied. The strain was applied in a step-like manner, first along the $x$ direction and then along the $y$ direction (Fig.~\ref{fig:str}(a)). It has been established that the application of $\varepsilon_x$ leads to an oppositely directed change in the measurable resistances $R_x$ and $R_y$ (Fig.~\ref{fig:str}(b)). Namely, the resistance $R_{x}$ along the strain axis $x$ increased, while $R_y$ measured along the perpendicular axis $y$ decreased (first two steps in Fig.~\ref{fig:str}). Further application of transverse strain results in similar changes along the perpendicular direction (the third and fourth steps in Fig.~\ref{fig:str}). Finally, the resistances along both axes became equal to each other (with an accuracy of 0.3\%) but were slightly above the initial value. Similar behavior is observed across the entire temperature range studied, except for the NCCDW-CCDW transition region.

\begin{figure}
\includegraphics[width=\linewidth]{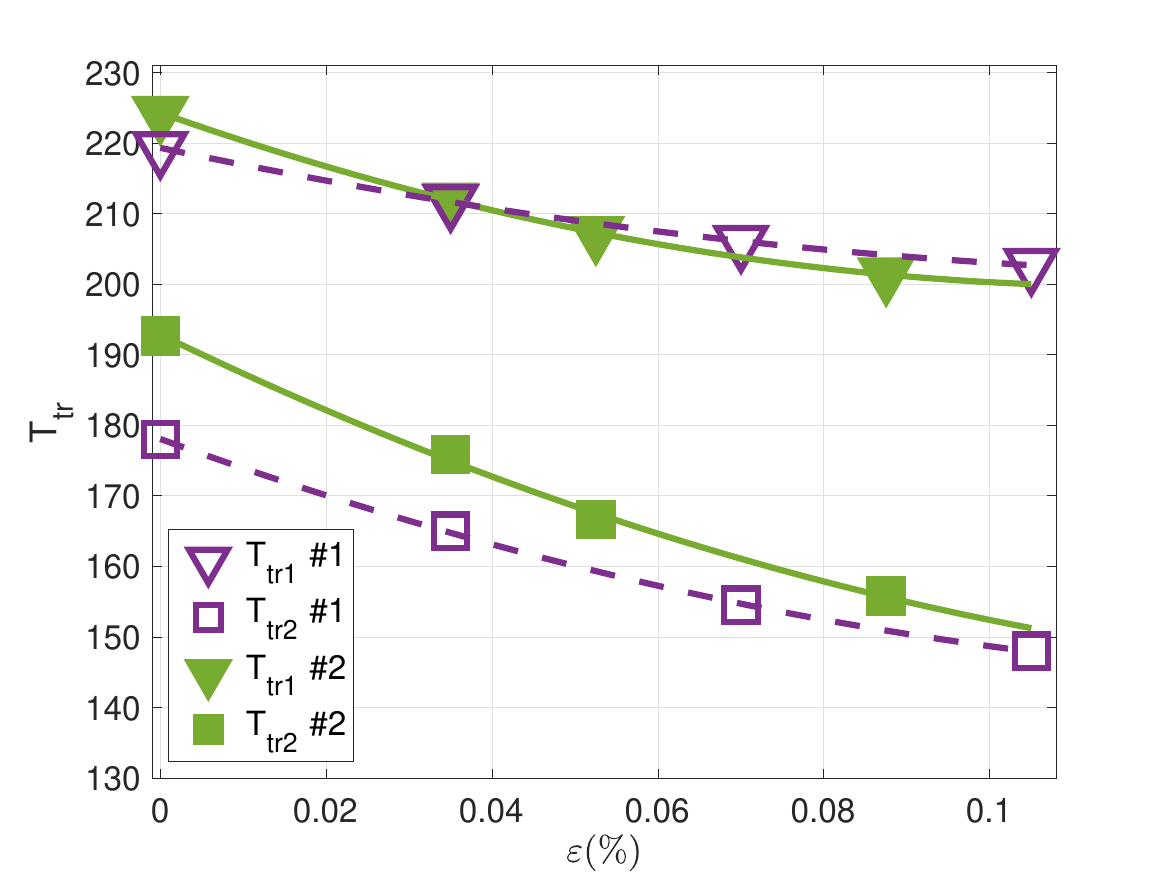}
\caption{\label{fig:deltaT} Dependence of transition temperatures in 1T-TaS$_2$ on the nominal biaxial strain of the substrate for samples \# 1 and \# 2.}
\end{figure}

Consider now the effect of biaxial strain on the temperature of the transition between the CCDW and NCCDW phases. Fig. \ref{fig:deltaT} shows the dependence of the change in transition temperatures from the CCDW phase to the NCCDW ($T_{tr1}$) and from the NCCDW phase to the CCDW ($T_{tr2}$) (Fig. \ref{fig:R(T)}) on the nominal strain of the substrate. Note that, under biaxial strain, both transition temperatures decrease. Moreover, $T_{tr2}$ decreases faster, leading to an expansion of the hysteresis loop. Note also that the relationship is not linear. As the strain increases, the sensitivity of the transition temperatures decreases. Thus, the application of biaxial tensile strain results in a decrease in the transition temperature and an increase in the hysteresis loop width.

\begin{figure}
\includegraphics[width=\linewidth]{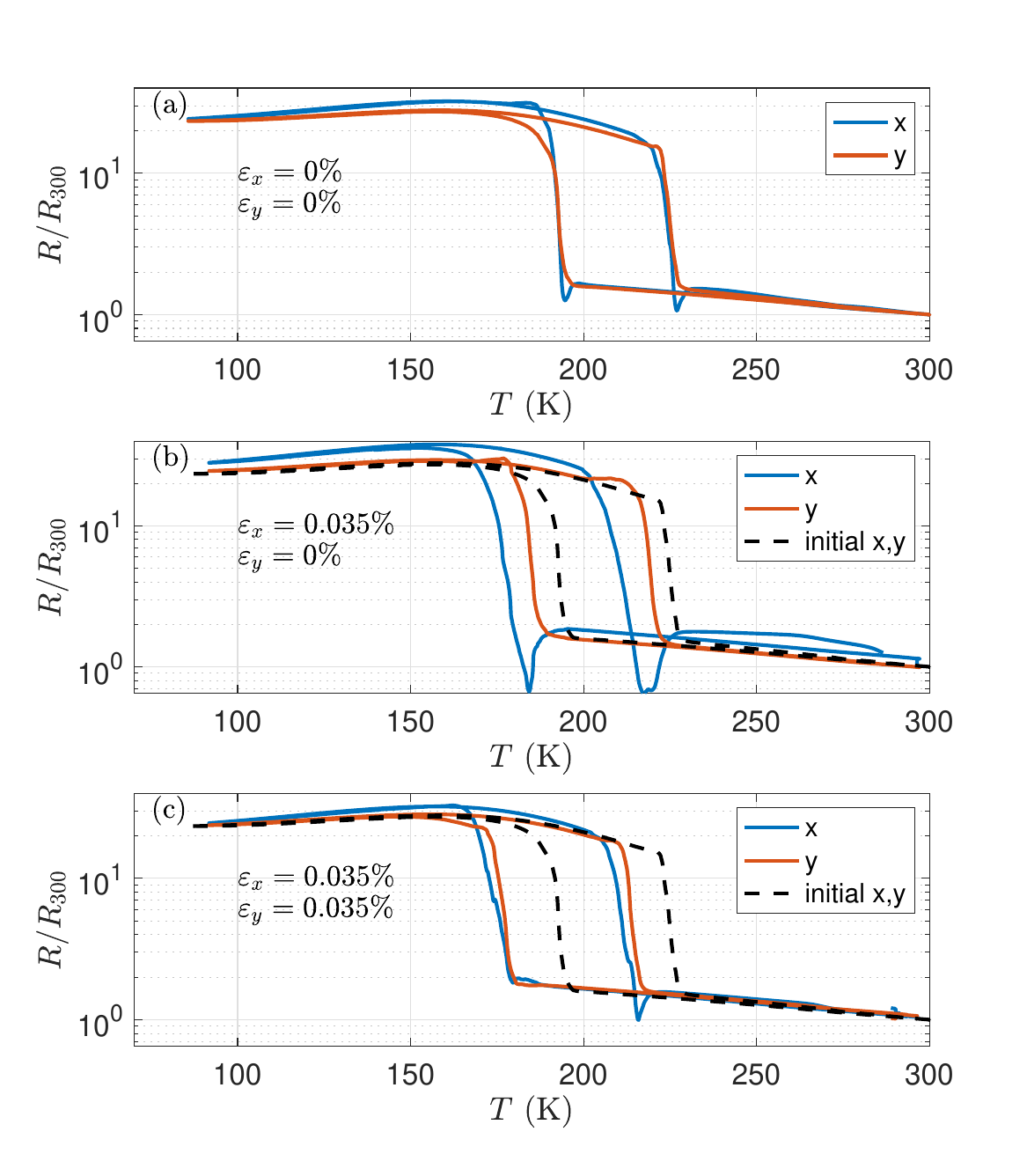}
\caption{\label{fig:RTstr} Temperature dependencies of the resistances $R_x$, $R_y$ measured in sample \# 2 along orthogonal directions under various strains: a) relaxed sample, b) uniaxial strain, c) biaxial strain, 
The sample and axes are shown in Fig.\ref{fig:samp}.}
\end{figure}

The splitting of the phase transition is illustrated by strain-dependent resistance measurements (Fig.~\ref{fig:RTstr}). In the initial relaxed state, the transition temperatures for $R_{x}(T)$ and $R_{y}(T)$ coincide (Fig.~\ref{fig:RTstr}(a)). Applying uniaxial strain along the $x$-axis shifts transition temperatures differently for the two directions (Fig.~\ref{fig:RTstr} (b)): the resistance along the strained direction ($x$) changes at a lower temperature than the perpendicular one ($y$). The splitting is substantial, reaching 10~K in this sample --- far beyond any experimental uncertainty. Notably, applying additional equal strain along the 
$y$-axis produces pure biaxial strain and practically eliminates the splitting (Fig.~\ref{fig:RTstr} (c)). 

The drop of the resistance $R_x$ near the splitting temperature region  (Fig.~\ref{fig:RTstr} (b)) is clearly seen.  This drop in $R_x$ is a consequence of an increase of $R_y$ (see Discussion for details). 

Thus, in the temperature range where the phase transition has occurred along one axis but not yet along another one, a pronounced conductivity anisotropy is observed. In our experiment, this anisotropic regime spans up to 20 K, with a maximum anisotropy resistance ratio of 20, which corresponds to the anisotropy of resistivity $\rho_{yy}/\rho_{xx}\gtrsim2$ (see below).

\section{\label{sec:level1}Discussion}

Let us start our discussion of the results by analyzing the effect of deformation on the measured resistance. It can be shown by using the equations of Ref.~\cite{miccoli2015100th} that for a 2D anisotropic rectangular sample with $\rho_{xx}$ and $\rho_{yy}$ as sheet resistances along $x$ and $y$ axes, the resistances $R_x$ and $R_y$ along $x$ and $y$ axes can be expressed as 

\begin{equation}
R_x=\frac{8}{\pi}\frac{\sqrt{\rho_{xx}\rho_{yy}}}{\sinh\left(\pi\frac{l_y\sqrt{\rho_{yy}}}{l_x\sqrt{\rho_{xx}}}\right)}
\label{eq1}
\end{equation}
and
\begin{equation}
R_y=\frac{8}{\pi} \frac{\sqrt{\rho_{xx}\rho_{yy}}}{\sinh\left(\pi\frac{l_x\sqrt{\rho_{xx}}}{l_y\sqrt{\rho_{yy}}}\right)},
\label{eq2}
\end{equation}
where $l_x$ and $l_y$ are the lengths of the rectangle along the $x$ and $y$ axes, respectively. One of the non-obvious consequences of the equations \ref{eq1} and \ref{eq2} is that an increase in the resistivity in one direction (e.g. $\rho_{xx}$) leads to a decrease in the resistance measured along another direction ($R_y$ in this case). This is what leads to the appearance of deep minima on $R_x(T)$ dependencies when $R_y(T)$ starts to increase upon cooling (Fig. \ref{fig:RTstr}(b)).

Using Eq. \ref{eq1}, \ref{eq2} and the resistance ratio $R_y/R_x=2.04$ of the relaxed sample \#1 shown in Fig.~\ref{fig:str}, we obtain the sample \#1 aspect ratio value $l_y/l_x=1.1194$. Then, taking into account that $\Delta l_{x,y}/l_{x,y}\ll \Delta R_{x,y}/R_{x,y}$, the strain-induced values corresponding to $\varepsilon_x=0.035 \%$ and $\varepsilon_y=0$ calculated by using Eq.~\ref{eq1},\ref{eq2} are $\rho_{xx}=1.0286\cdot\rho_0$ and $\rho_{yy} =1.0008 \cdot\rho_0$, where $\rho_0$ is the resistivity of the relaxed sample. Doing the same calculations for sample \# 2 with $R_y/R_x=36.94$, we obtain $l_y/l_x=1.72$, $\rho_{xx}=1.0382\cdot\rho_0$ and $\rho_{yy} =1.0008 \cdot\rho_0$. Both results are surprising, as due to the non-zero Poisson's ratio $\nu$ ($\nu\approx 0.4$ \cite{bauer1989_Poisson}), a strain of the substrate in the $x$ direction $\epsilon_{xx}$ is accompanied by its shrinkage in the perpendicular direction $-\nu \epsilon_{yy}$, where $\epsilon_{xx}$ and $\epsilon_{yy}$ are components of the deformation tensor. Naively, we expect in this case $\Delta\rho_{yy}=-0.4\Delta\rho_{xx}$, whereas in practice $\Delta\rho_{yy}$ and $\Delta\rho_{xx}$ are both positive and $\Delta\rho_{yy}\ll\Delta\rho_{xx}$. 

To explain this unusual result, we must suggest that the tenzoresistivity tensor $\Pi$ is 
non-diagonal, $\pi_{xy}\neq 0$. In this case, the effect of biaxial strain on the resistivity tensor must be written as
\begin{equation}
\frac{1}{\rho_0}
\begin{pmatrix}
\Delta\rho_{xx}\\ \Delta\rho_{yy} 
\end{pmatrix}=
\begin{pmatrix} 
\pi_{xx}&\pi_{xy}\\ 
\pi_{yx} &\pi_{yy} 
\end{pmatrix}
\begin{pmatrix}
\epsilon_{xx}\\
-\nu\epsilon_{xx} \end{pmatrix},
\label{eq:matrix}
\end{equation}
where $\pi_{xx}=\pi_{yy}$ and $\pi_{yx}=\pi_{xy}$. 
Substituting $\epsilon_{xx}=3.5\cdot 10^{-4}$, $\nu =0.4$, $\Delta \rho_{xx}/\rho_0=0.0286$ and $\Delta\rho_{yy}/\rho_0=0.8\cdot 10^{-3}$ to Eq.~\ref{eq:matrix} for the first sample, one gets
\begin{equation}
\Pi_1=\begin{pmatrix}98.4&41.6\\
41.6&98.4\end{pmatrix}.
\label{eq:pi1}
\end{equation}
Similarly, for the second sample,  
\begin{equation}
\Pi_2=\begin{pmatrix}131.0&54.7\\
54.7&131.0\end{pmatrix}.
\label{eq:pi2}
\end{equation}
The results are remarkably consistent, especially considering the difference between the experimental and theoretical geometries. 

The non-zero off-diagonal component in the response matrix is a feature of the NCCDW state of 1T-TaS$_2$. In 1T-TaS$_2$, the NCCDW  consists of domains of a CCDW phase containing several tens of David stars and separated by domain boundaries. Each domain boundary removes approximately 1/3 of the period of the commensurate structure \cite{wu1990}. Therefore, $ q_{NCCDW} >q_{CCDW}$, in agreement with XRD data \cite{scruby1975role}. The mean distance between domain boundaries $d$ is proportional to the difference between the actual CDW wave vector and its commensurate values, 
$\delta q=q_{NCCDW}(T)- q_{CCDW}$, $d=2\pi/3\delta q$. As the CDW gap is suppressed in the domain boundary regions, their conductivity is much higher than that of the intra-domain region, which has a 0.4~eV low-temperature energy gap \cite{ma2016metallic}: the latter provides only $\sim 3-5$\%\ of the total conductivity in the NCCDW state (see Fig.~\ref{fig:R(T)}). Thus, the conductivity of the domain boundary network is determined by both the intrinsic conductivity of each boundary and their spatial density. Therefore, applying uniaxial strain $\epsilon_{xx}$ to a sample changes its properties in two ways: it modulates $\rho_{xx}$ and alters the mean distance $d_x$ between the boundaries aligned perpendicular to the strain direction. It is for this reason that the non-diagonal matrix elements $\pi_{xy}=\pi_{yx}$ are not equal to zero. The positive sign of $\pi_{xy}$ means $dd_x/d\epsilon_{xx}>0$ (increase in the domain size in $x$ direction with the strain $\epsilon_{xx}>0$), which leads to $d\delta q/dd_{x}<0$. In other words, uniaxial strain brings the CDW closer to commensurability in the direction of the strain.

\begin{figure*}
\vskip4mm
\includegraphics[width=0.245\linewidth]{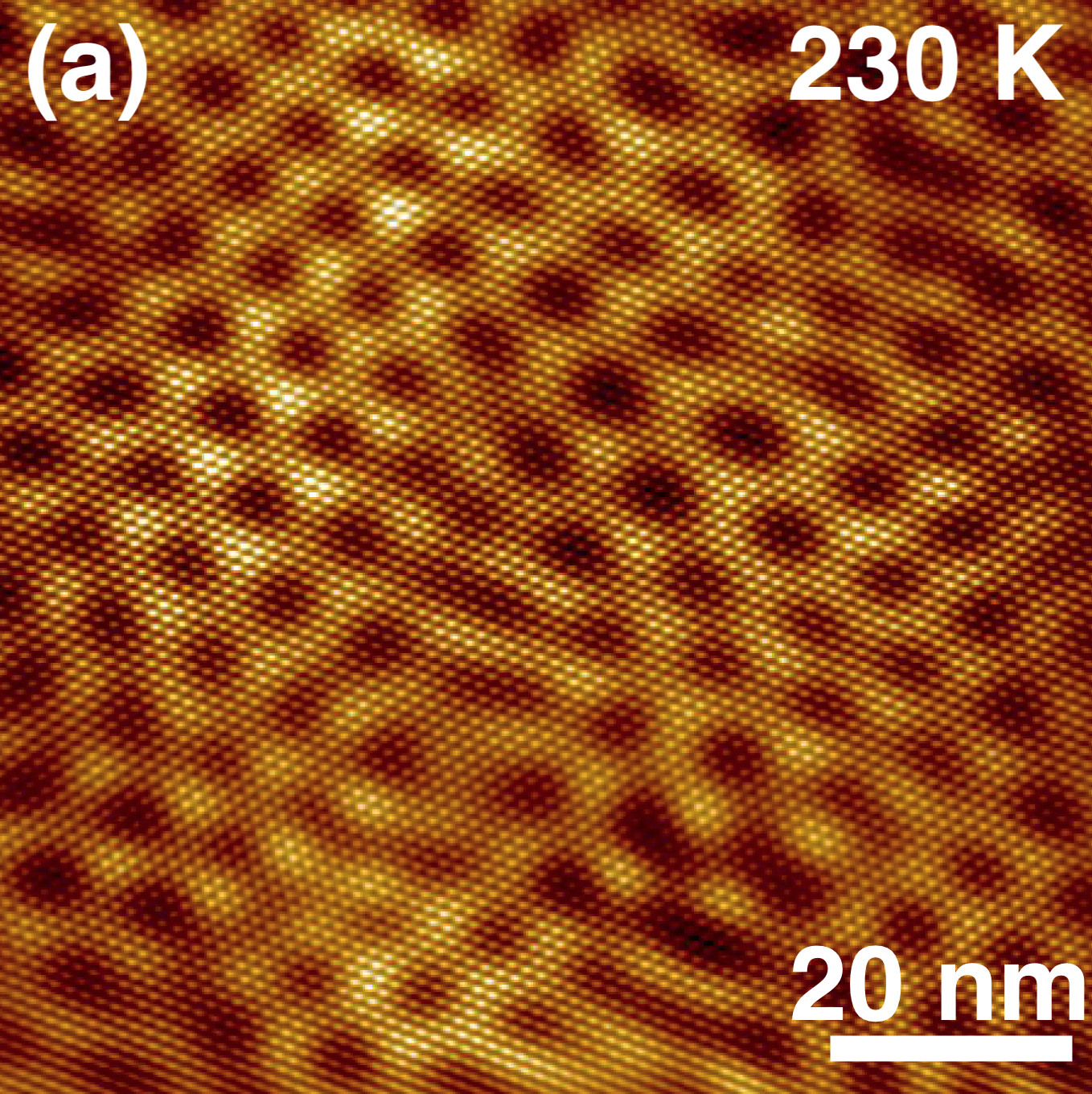} \includegraphics[width=0.245\linewidth]{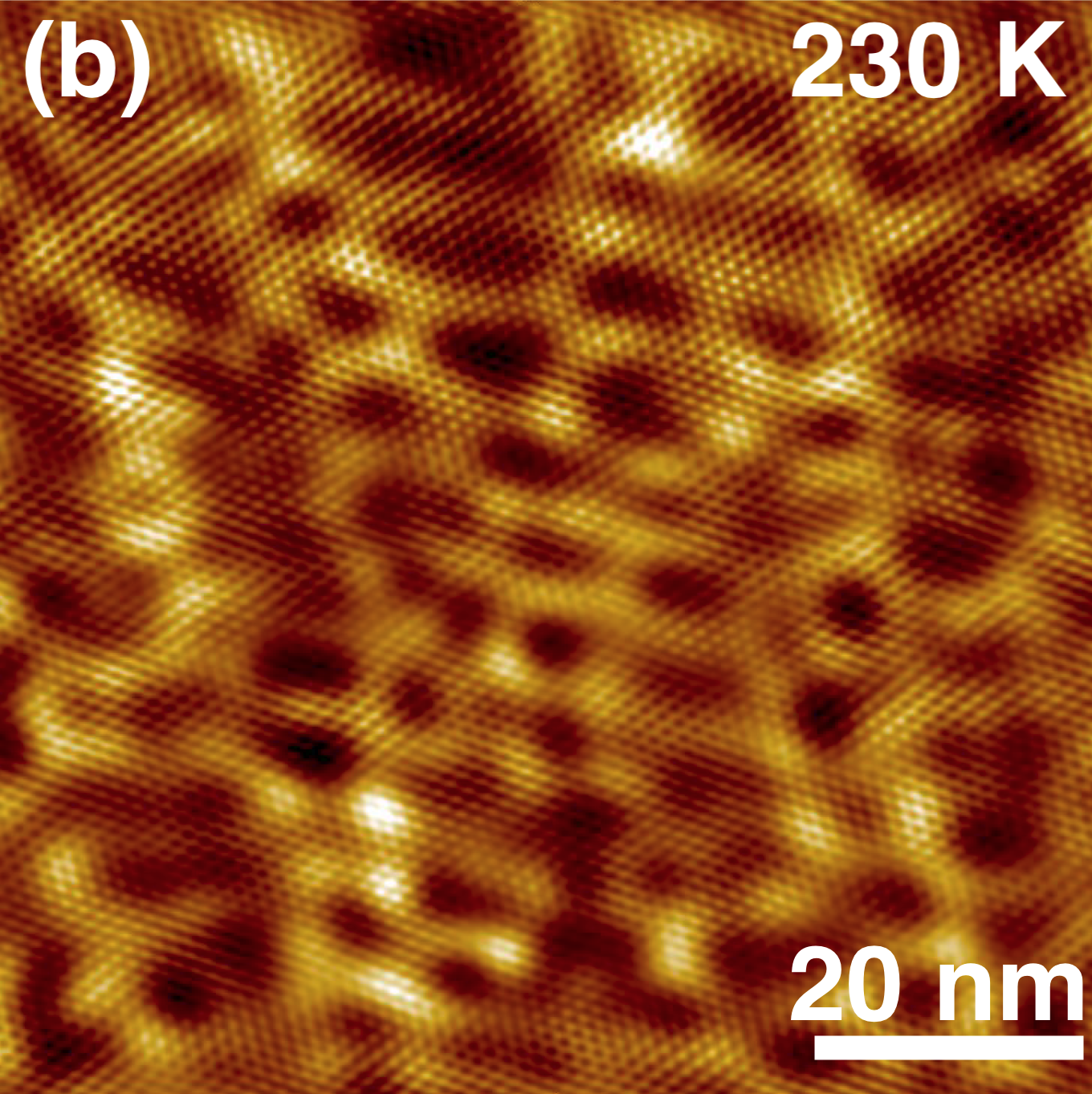} \includegraphics[width=0.245\linewidth]{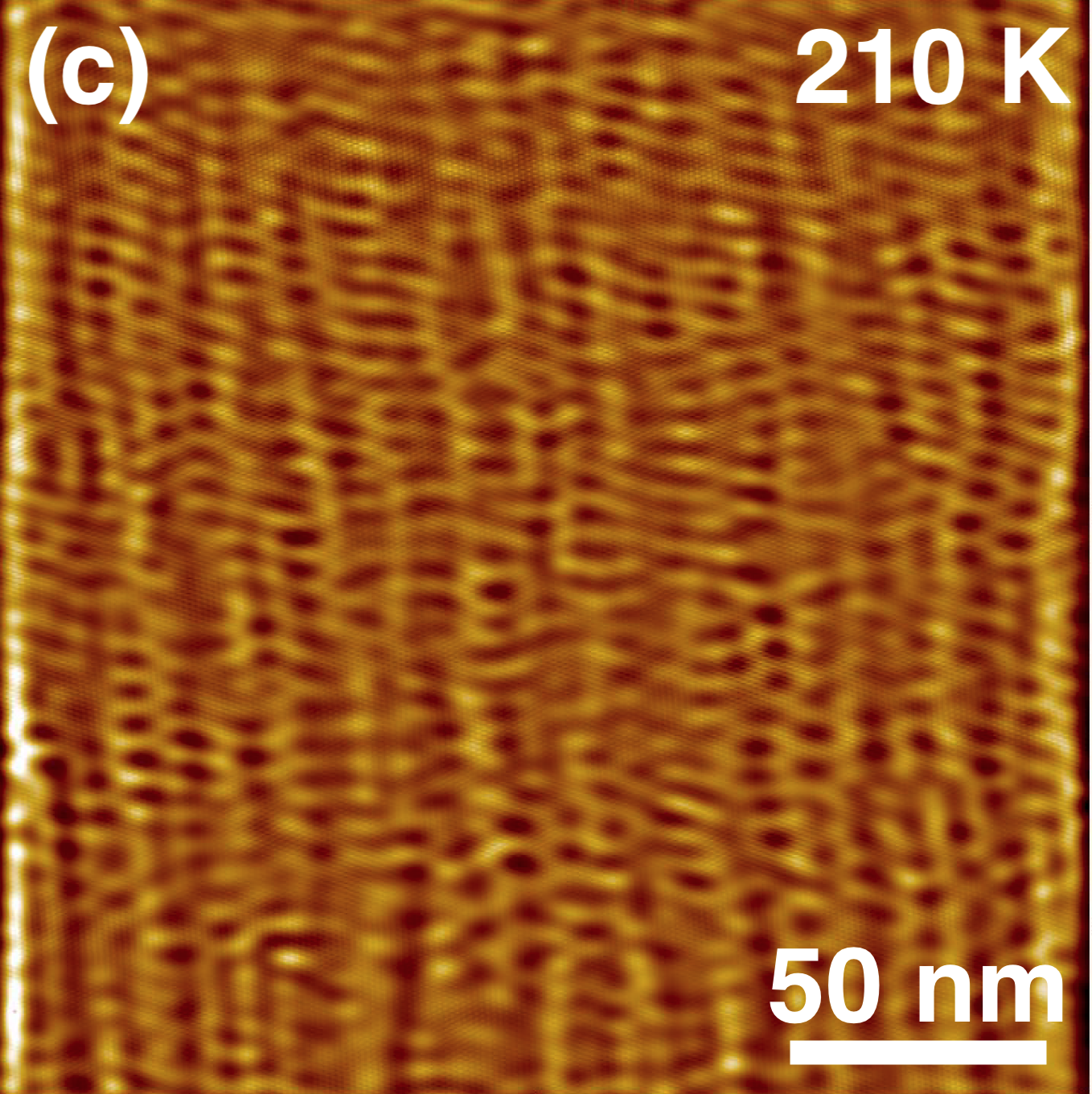} \includegraphics[width=0.245\linewidth]{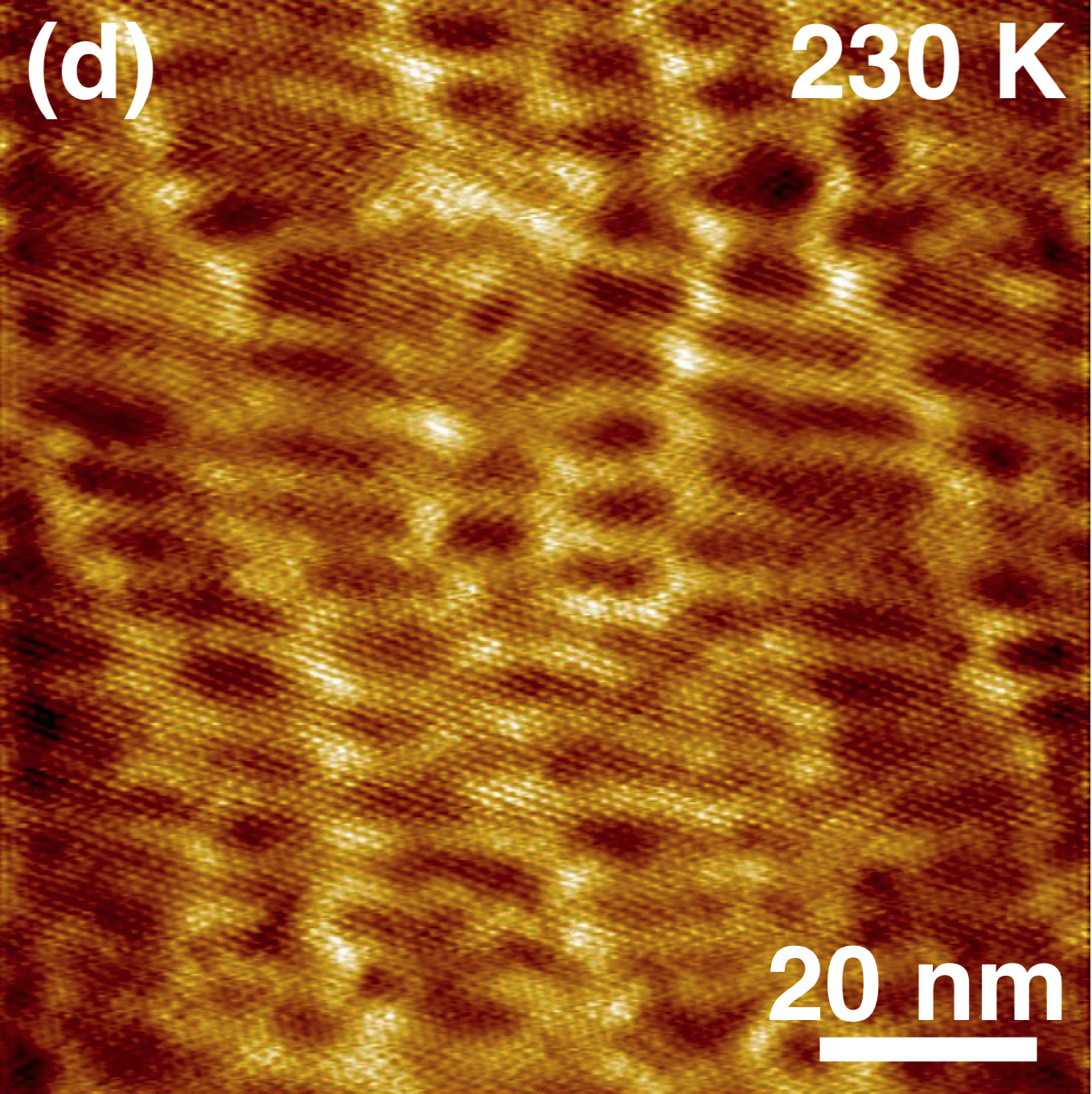}   
\caption{\label{fig:net} Typical STM images of 1T-TaS$_2$ surface under uniaxial strain. (a) Relaxed state; (b) nominal deformation $\epsilon_{xx}=0.4$\%; (c) nominal deformation $\epsilon_{xx}=$0.4\%; d) nominal deformation $\epsilon_{xx}=1.3$\% . All the images were obtained after cooling from room temperature. Set point: $V_t=0.3$~V, $I=0.1$~nA. } 
\end{figure*}

The above is confirmed by the results of STM measurements. Figure~\ref{fig:net} shows the evolution of the superstructure that forms in the temperature region of the CCDW–NCCDW transition ($T=230$~K) on the surface of 1T-TaS$_2$ under uniaxial stretching of the crystal. In the relaxed crystal, \ref{fig:net}(a), a slightly distorted hexagonal [superstructure] is observed, similar to the well-known high-temperature one. Stretching the crystal along the x-axis leads to the appearance of disordered domains elongated along the stretching axis (\ref{fig:net}(b)).
 As the temperature decreases (\ref{fig:net}(c)) to 210 K where anisotropy reaches its maximum (see Fig.~\ref{fig:RTstr}(b)), a quasi-periodic structure consisting of domains elongated along the stretching axis is formed. With further increase of deformation (Fig.~\ref{fig:net}(d)), the domain length along the x-axis increases noticeably, while their transverse size slightly decreases further. 

The anisotropy of the structures shown in Fig.~\ref{fig:net}(d) is obvious: the domain length in the stretching direction is 2–5 times greater than its transverse dimensions. In such a deformed state, the crystal contains almost the same number of horizontal conduction channels, but substantially fewer vertical ones. Note that, with increasing stretching, domain boundaries also change—a small increase in the brightness and width is noticeable in Fig.~\ref{fig:net}(d). A priori, it is not obvious exactly how such changes in the period and width of domain boundaries oriented along and across the stretching direction will affect the components of the resistivity tensor. $R(T)$ data in Fig.~\ref{fig:RTstr} shows that such changes are still not big. 

An increase in the domain size along the stretching direction means an approach to commensurability. Therefore, transitions along the stretching axis (x-axis) occur at a higher temperature than in the transverse direction, which leads to splitting of the CCDW–NCCDW phase transition. If so, the resistivity along the $x$ direction remains almost the same, but along the $y$ direction, it increases significantly owing to singularity in the domain length in the strain direction. This is exactly the situation observed: in a narrow temperature range, $R_y(T)$ already jumps into a commensurate state, but $R_x(T)$ is still in a nearly commensurate state (Fig.~\ref{fig:RTstr}(b)). 

In the present study, we did not succeed in STM observation of splitting  of the transition with our bending setup -- at temperatures below 200 K highly nonuniform structures consisting of a mixture of chaotically distributed low-temperature ordered islands of different phases are formed. We attribute this behavior with clearly observed slow tip-induced evolution of nonequilibrium structures.

The observed suppression of the CDW-NCCDW transition results from a slight strain-induced modification of the band structure and CDW wave vectors. The experimental results allow us to estimate the value of $dq/d\epsilon$. In 1T-TaS$_2$, $q_{NCCDW}$ increases nearly linearly with temperature from 0.282 at $T=250$ K to 0.285 at $T=303$~K \cite{scruby1975role}. Therefore, $dq/dT=5.7\cdot 10^{-5}$ and $dq/d\epsilon=dq/dT\cdot dT_p/d\epsilon=-3.2$. It is the dependence of the CDW wave vector on the deformation that leads to a decrease in the NCCDW-CCDW transition temperature. For comparison, in the quasi-1D conductor {\it o}-TaS$_3$ $dq/d\epsilon = 0.17$-0.33 \cite{zybtsev2016quantization} that is one order smaller.

To the best of our knowledge, the observed splitting of a phase transition is a unique phenomenon that is absent in systems of other types (superconductors, ferro- and antiferromagnetic, {\it ect.}). The possibility of splitting the CCDW-NCCDW phase transition arises from the presence of three components of the order parameter and the ability to manipulate their relationships to each other by the application of a uniaxial strain. We would like to note that there is a 2D system where two CDWs form at different temperatures due to a small initial distortion of the lattice, namely RTe$_3$, where R is a rare-earth metal \cite{dimasi1995chemical}. 
In this system, the Cmcm symmetry crystal lattice with an almost square rectangular base ($a\approx b$) also allows for switching the direction of the CDW by unidirectional strain \cite{gallo_Nat_2024}. Surprisingly, the CDW amplitude is noticeably suppressed in this system when $a=b$ under unidirectional strain.

In conclusion, we have studied the effect of uniaxial and biaxial strain on the resistivity of 1T-TaS$_2$. A unique feature was discovered: a strain-induced splitting of the phase transition.
The splitting of the resistive CCDW-NCCDW phase transition is explained by a change in the shape and size of the domains of the NCCDW when uniaxial strain approaches commensurability with the CDW wave vector along the strain direction. Not less surprising is the effect of uniaxial strain on the resistivity tensor components at the room temperature. Instead of being a simple diagonal matrix with possibly negligible off-diagonal components, the tenzoresistivity tensor has compatible values of diagonal and off-diagonal components ($\pi_{xx}/\pi_{xy}\approx 0.4$, $\pi_{xx}=\pi_{yy}$, $\pi_{xy}=\pi_{yx}$). This relationship results from the dominant contribution of the domain wall network to conductivity, whose strain-induced deformation leads to a large transverse contribution to the resistivity changes. 
Our results reframe the CCDW–NCCDW transition in 1T-TaS$_2$ as a directionally tunable, domain-network-controlled phenomenon.

The data that support the findings of this study are available from the corresponding author upon reasonable request.

\bibliography{apssamp}

\end{document}